\documentclass[%
 reprint,
 amsmath,amssymb,
 aps,
floatfix,
]{revtex4-2}

\usepackage{graphicx}
\usepackage{dcolumn}
\usepackage{bm}

\usepackage[x11names]{xcolor}
\usepackage{framed}
\usepackage{quoting}

\usepackage{appendix}
\usepackage{lipsum}

\definecolor{beaublue}{rgb}{0.74, 0.83, 0.9}
\definecolor{cerulean}{rgb}{0.0, 0.48, 0.65}
\definecolor{frenchblue}{rgb}{0.0, 0.45, 0.73}

 \colorlet{shadecolor}{beaublue}

\begin{document}

\preprint{APS/123-QED}

\title{The simple emergence of complex molecular function}

\author{Susanna Manrubia}
\affiliation{Department of Systems Biology, National Centre for Biotechnology (CSIC). \\ c/ Darwin 3, 28049 Madrid, Spain \\
Interdisciplinary Group of Complex Systems (GISC), Madrid, Spain}%

\date{\today}

\begin{abstract}
At odds with a traditional view of molecular evolution that seeks a descent-with-modification relationship between functional sequences, new functions can emerge {\it de novo} with relative ease. At early times of molecular evolution, random polymers could have sufficed for the appearance of incipient chemical activity, while the cellular environment harbors a myriad of proto-functional molecules. The emergence of function is facilitated by several mechanisms intrinsic to molecular organization, such as redundant mapping of sequences into structures, phenotypic plasticity, modularity, or cooperative associations between genomic sequences. It is the availability of niches in the molecular ecology that filters new potentially functional proposals. New phenotypes and subsequent levels of molecular complexity could be attained through combinatorial explorations of currently available molecular variants. Natural selection does the rest.  
\end{abstract}

\maketitle

\section{Introduction}

Half a century ago, the idea that gene specificity could rely on a unique protein sequence raised concerns regarding the come into being of functional genes. Natural selection would be ineffective if the raw material on which it had to act were random sequences, given that a myriad of universes as old and as large as ours, where random sequences would be systematically generated, appeared absolutely insufficient to produce the tiniest functional molecule \cite{salisbury:1969}. The apparent paradox, whose assumptions echoed intelligent design arguments \cite{ogbunugafor:2020}, was obviating the overwhelming number of neutral and quasi-neutral mutations \cite{kimura:1971,ohta:1973,maynard-smith:1970} and the fact that partial functionality is better than no functionality at all \cite{dawkins:1996}. Actually, molecular complexity might be relatively straight to achieve.

Molecular function is highly redundant. The same molecular phenotype can be obtained from an astronomically large number of different genotypes, as revealed by computational and empirical studies of complete genome spaces for short sequences \cite{louis:2016,ahnert:2017,garcia-martin:2018}. Such genotypes are organized as networks-of-networks \cite{yubero:2017,aguirre:2018}, a non-trivial topology with important dynamic consequences \cite{aguirre:2018} that also points to the need of an updated metaphor to represent adaptive landscapes \cite{catalan:2017}. The abundance of phenotypes is not homogeneous in sequence space \cite{wagner:2011,dingle:2015,ahnert:2017,manrubia:2017}, since most sequences are mapped into a small fraction of very large phenotypes. This fraction, however, suffices to guarantee a spectrum of different and efficient enough functions so as to sustain life as we know it. The high dimensionality of genotype spaces and the fact that the networks of abundant phenotypes percolate the space of sequences under very general conditions \cite{gavrilets_JTB:1997} further ensures that different functions may be awaiting just a few mutations apart \cite{fontana:2002,schultes:2000}. 

The genotype-to-function map is redundant in several different ways beyond neutral and quasi-neutral mutations. The many layers of expression from genotype to function \cite{manrubia_PLR:2021} act towards increasing the ensemble of possible genotypes coding for comparable phenotypes \cite{arias:2014,catalan:2018}. Further, function is flexible, so phenotypes admit a range of variation, and phenotypes are plastic, so their expression adapts to different environments \cite{scott:2012,sommer:2020}. Beyond multiple inconsequential variations in genotypes, also changes in molecular structure or composition might be irrelevant for the functionality of a phenotype. 

Molecular mimicry \cite{wucherpfennig:2001}, protein moonlighting \cite{piatigorsky:2007,singh:2020} or enzyme promiscuity \cite{khersonsky:2010} are three widespread expressions of phenotypic redundancy and functional flexibility. Molecular mimicry was originally defined in the context of immunology: occasionally, self-peptides are sufficiently similar in sequence so as to be mistaken by pathogen-derived peptides and, consequently, trigger an autoimmune response \cite{wucherpfennig:2001}. The concept, however, can be extended to embrace the many instances where molecules of dissimilar composition disguise to resemble others' structure \cite{tsonis:2008}. In protein moonlighting, the same protein may have multiple, context-dependent functions; this property has been also reported for RNA molecules \cite{vaidya:2012}. Enzyme promiscuity is a conceptually related notion that refers to the ability of an enzyme to catalyze reactions other than those for which it was in principle selected. RNA promiscuity naturally appears when structures compatible with a given sequence, but different from the minimum-free-energy secondary structure, are considered \cite{wagner:2014}. The former mechanisms illustrate complementary properties of relevance in evolution: on the one hand, molecules of different origins might perform similar functions; on the other hand, the same molecule can be recruited to perform a different function. Phenotype tolerance to endogenous and exogenous variation implies that genes and, in general, any sort of functional molecule may not need initial adjustments to engage into a secondary function \cite{conant:2008}. But, once in place, natural selection can act towards optimization, if needed. 

The skewed distribution of phenotype sizes conditions what is visible to evolution \cite{schaper:2014} but may, under certain circumstances, also facilitate the appearance of simple molecular functions {\it de novo}. The emergence of function has in all likelihood been relevant in the origin of an RNA world \cite{briones:2009,wachowius:2021} and in the genesis of simple replicators \cite{catalan:2019a}. Still, single gene or single molecule redundancy, and the selective improvement of their function through point mutations, only represents the fine-tuning of molecular evolution. Once originated and optimized, small functional sequences might act as the basic bricks of multi-purpose molecules \cite{manrubia:2007} through a modular constructive principle that applies from proteins \cite{trifonov:2009} to organisms \cite{ryan:2006,wagner-gp:2007}. 

Ensembles of agents that replicate independently ---which, in a certain sense, may act as competitors--- can integrate to form a new, more complex entity \cite{west:2015} through what is known as a major evolutionary transition \cite{maynard-smith:1995}. Many of these transitions are cooperative \cite{stewart:2020}, a paradigmatic example at the molecular level being the emergence of chromosomes. However, there are multiple examples, especially in the viral world, where flexible cooperation, that is, without irreversible fusion of the parts, appears as a successful adaptive strategy. In viral quasispecies, ecological roles can be allocated in different mutant classes \cite{colizzi:2014}, and collective cooperation may be needed to maintain pathogenesis \cite{vignuzzi:2006}. In viruses, horizontal gene transfer (HGT) is not only a common mechanism for adaptation, but probably also a way to generate new viral species \cite{iranzo:2016b}. A remarkable form of distributed cooperation is that of multipartite viruses, whose genes are propagated in independent capsids \cite{sicard:2016,lucia-sanz:2017}. Viruses are a powerful system of generation of new molecular function. Together with other mobile elements, they function as transporters of genomic sequences and may assist their integration in higher organisms. In the coevolution of viruses with hosts, the former may even promote increases in host complexity \cite{seoane:2020} and spur major evolutionary transitions \cite{koonin:2016}. 

In the forthcoming sections, we will focus on specific examples that illustrate how some of the mechanistic principles outlined in this introduction can be used to devise scenarios where complex molecular function could parsimoniously emerge. The conceptual framework integrates low-cost pieces with incipient functionality, different cooperative schemes and an ecological context (or an organized molecular environment) responsible for the selective pressures that act on such incipiently functional systems. Our aim is to link robust features of the molecular genotype-to-structure map, contingency, and selective evolution, and to discuss how, at the next level, innovation can emerge through distributed cooperation. 

\section{Phenotypic bias: the free lunch of molecular function}

RNA has been studied in depth as a paradigmatic example of the sequence-to-structure map \cite{schuster:1994}. It has been shown that the frequency distribution of secondary structures follows a log-normal distribution \cite{dingle:2015,manrubia:2017}, thus quantifying the bias in phenotype (structure) abundance. Accordingly, common phenotypes are orders of magnitude more frequent than average-sized or small phenotypes (which are, by any reasonable measure, invisible to evolution). RNA structure is tightly related to molecular function and, as such, it may have played a main role at the early stages of chemical evolution and, especially, in an RNA world predating modern cells \cite{gilbert:1986}. The repertoire of secondary structures in large populations of short RNA polymers is limited \cite{gruner:1996}: topologically simple RNA modules are abundant. In open RNA chains, there is a predominance of stem–loops (composed of a stem and a hairpin loop) and hairpin structures \cite{stich:2008}, while closed RNA chains preferentially fold into rod-like structures (a stem closed by two hairpin loops) \cite{cuesta:2017}. This fact is solely based on thermodynamic principles \cite{lorenz:2011} and implies that, by default, short RNA sequences will predominantly yield a handful of structures.

The relevant implication of this high redundancy is that some of those abundant structures might be the end result of random RNA polymerization (as it could have been the case in prebiotic environments \cite{huang:2003}). With no previous selection, random polymers might have thus covered an array of incipient functions. Such is the case of a variety of hairpin-like structures able to promote ligation reactions \cite{buzayan:1986,fedor:1999}, or of hammerhead structures involved in cleavage \cite{delapena:2017-M}. RNA self-ligation might indeed be instrumental in the modular construction of more complex ribozymes, as theoretically proposed \cite{briones:2009} and empirically shown \cite{gwiazda:2012,staroseletz:2018}.

There are reasons to believe that the severe phenotypic bias of the sequence-to-structure RNA map, as quantified through the log-normal distribution of phenotype sizes, is a general property of genotype-to-phenotype maps \cite{manrubia:2017,catalan:2018,garcia-martin:2018,manrubia_PLR:2021}. If this is the case, the scenario described for RNA should hold broadly, and apply to other polynucleotides, to peptides, and to polymers at large. The next step in the construction of chemical complexity regards the emergence of cooperative behaviour of some kind \cite{villarreal:2021}, such that the proto-functionalities provided by such molecules do not get lost. Eigen's hypercycles \cite{eigen:1978} were an early proposal in this respect that has been deeply explored \cite{szostak:2016}. Indeed, some experiments indicate that networks of interacting molecules may have formed early in chemical evolution, pointing also at an intrinsic ability of RNA to evolve complexity through cooperation \cite{vaidya:2009}. 

\section{Emergence of viroid-like replicators}

Genotype-to-phenotype redundancy is an intrinsic property of natural systems \cite{jimenez:2013,payne:2014b}. This redundancy should have been important in the prebiotic origin of chemical function, but also all along evolution: as first enunciated in geology \cite{hutton:1785}, chemical processes should have acted in the same manner and with essentially the same intensity in the past as they do in the present. We must add that the effect of such processes is different because it is the molecular ecosystem that has changed. For instance, it seems reasonable to assume that hairpin-like RNAs with potential ligase activity are continuously produced in the RNA-rich cellular environment \cite{maurel:2019}, where multiple RNA sequences of various origins might be present: fragments of RNA transcription \cite{tuck:2011}, transient products of RNA degradation \cite{houseley:2009}, pieces of viral genomes \cite{combe:2015} or rod-shaped microRNAs \cite{bartel:2004,he:2004}. Natural processes should be inadvertently and steadily generating raw proto-functional sequences. However, such proposals will be filtered by purifying selection, by competition with (probably fitter) existing functions or simply by degradation so, at odds with what happens in geological systems, they mostly disappear without a trace. With an exception: when an ecological niche is available. 

Viroids are small, non-coding, circular RNA molecules~\cite{diener:1971}. 
Despite having a small genome of a few hundred nucleotides, they behave as competent and persistent replicators in higher plants, apparently their only natural hosts. The evolutionary origin of viroids has been a matter of discussion, and various hypotheses have been put forward \cite{diener:1996,diserio:2017}. It has been suggested that viroids may be related to other extant cellular RNAs \cite{diener:1981}, could have originated from retroelements or retroviruses \cite{kiefer:1983}, or be ancient relics of a precellular RNA World~\cite{diener:1974,flores:2014}. All these possibilities seek the origin of viroids in a previously functional system, assuming shared ancestry and a broad phylogeny linking viroids to other extant functional RNAs \cite{diener:1991,elena:1991}. 
This assumption has been criticized on the basis that the observed sequence similarity might be spurious \cite{jenkins:2000} (see also referee reports in \cite{diener:2016}), attending as well to the high mutation rates of viroids \cite{gago:2009}, the diversity of their populations \cite{codoner:2006} and their low sequence conservation \cite{glouzon:2014}.  

Actually, it cannot be discarded that modern viroids emerged once eukaryotic cells had evolved \cite{zimmern:1982,catalan:2019a}. Inspection of the two families of viroids reveals important differences in their structure and replication cycle, as well as in the interactions with host proteins \cite{ding:2009,palukaitis:2014}. The two families are so different that a polyphyletic origin cannot be ruled out \cite{jenkins:2000}. In a context of a {\it de novo} emergence of rudimentary replicons, phenotypic bias may have played a role: for example, all circular RNAs of length 20 and lower fold into rod-like structures \cite{cuesta:2017,catalan:2019a}. Remarkably, the rod-like structure of viroids could be a case of molecular mimicry, since that structure resembles dsDNA and facilitates the recognition by RNA polymerases \cite{flores:2014}. Hence, a small rod could be "recognized" by the replicating machinery of the cell, triggering its replication and therefore its differential selection. 

Viroids exhibit a modular structure that has prompted their description as a ``collection of structural motifs which play specific functional roles in viroid replication, processing, transport, and pathogenesis''~\cite{steger:2016}. The first step in their emergence, as described in the previous paragraph, could have been relatively straightforward, as that of other potentially useful functions in the same or nearby environments. Once kick-started, their modularity strongly supports the possibility that different functional features of viroids could have been acquired through recombination with RNAs of different origins transiently present in the molecular ecology where they evolved. This is more than a conjecture, since some viroids are known to arise from other viroids through recombination \cite{hammond:1989,rezaian:1990}. A highly plausible case of modular recombination is provided by hepatitis $\delta$ virus \cite{chen:1986}, which has a viroid-like non-coding domain \cite{elena:1991} linked to a coding domain of independent evolutionary origin~\cite{weiner:1988}. 

Despite justified criticism of proposals that assume an unbroken phylogenetic connection between extant viroids and their potential ancestors in an RNA world, there is reason to believe that the molecular niche occupied by viroid-like molecules may have been continuously available (and likely occupied) given a minimal chemical complexity. It will be extremely difficult, if not impossible, to solve this question empirically. However, it comes to reason that any self-replicating system, however rudimentary, is subject to the emergence of parasites \cite{iranzo:2016d,koonin:2017}, of non-cooperative defectors that use system's resources for their sole benefit. The cellular environment is an extremely rich and varied ecology \cite{nathan:2014} where multiple control mechanisms, up to cell death \cite{vandoorn:2005} are acting and, indirectly, thus limiting the selfish escape of functional molecules. Another mechanism contributing to the preservation of system's integrity might be the difficulty of newcomers to invade a functional ecology \cite{brockhurst:2007}. Molecules that may potentially occupy a given niche may find it difficult to succeed if the system already has an optimized solution for that function. This sort of non-invasibility principle implies a degree of phylogenetic continuity, and endows first-comers with an implicit advantage. 
Extant viroids, as a possible example of such process at a molecular level, might be a combination of contingency (a frozen accident) and continuity as a side-effect of non-invasibility. In a viroid-free situation, it is highly likely that similar viroid-like replicators would emerge in short evolutionary time, occupying the vacant niche in the same way that radiating species do \cite{losos:2010}. Macroecological niches enjoy long periods of functional stasis, even in the face of taxonomic variability \cite{blanco:2021}. At odds with macroevolution and macroecology \cite{weber:2017}, an integration of molecular ecology and evolution (and phylogeny) is yet to be worked out. 

\section{Viral gene sharing}

Viruses are extremely abundant, diverse, strict molecular parasites that, perhaps with rare exceptions, infect all cellular organisms on Earth. It is difficult to overstate the role that viruses may have played in the construction of our complex biosphere \cite{witzany:2019,domingo:2020}. They are motors of biodiversity \cite{koonin:2006}, regulate ecosystems and global biogeochemical cycles \cite{weitz:2012}, and constitute a huge reservoir of genetic diversity \cite{suttle:2005}. Metagenomic techniques are enormously enlarging the quantity and quality of previously described viral species, and strongly suggest that we have only grasped the surface of viral gene diversity \cite{edwards:2005,zhang-ARV:2019}. 

Viruses mutate much faster than cellular genes \cite{sanjuan:2010}: a viral gene explores in a few thousand years a sequence space comparable to that explored by a typical nuclear gene since the Cambrian explosion \cite{abroi:2011}. Single viral populations are organized in viral quasispecies, swarms of mutants where each sequence may differ from each other in at least one mutation \cite{domingo:2019}, allowing a much more efficient exploration of sequence spaces \cite{villarreal:2008}. Not surprisingly, viral sequences found in metagenomic studies are dominated by rare genes, with up to 90\% of DNA reads encoding proteins not found in other cellular organisms, or in other viruses \cite{kristensen:2010}. The evolutionary freedom enjoyed by viruses may turn them into cradles of functional diversity. A substantial part of organismal evolution could be virus driven, since viruses contribute essential (functional) pieces that promote complexity increases in organisms \cite{villarreal:2021}. Viruses harbor protein domains with folds unknown in cellular organisms, and some of these domains have been transferred to the host \cite{abroi:2011}. Massive transfers of genes from virus to host are not uncommon \cite{liu:2010,gilbert:2017}. 

Viruses are extreme examples of mosaicism. They are truly chimeric in their composition, most often the emergent result of broad and wide viral gene sharing \cite{iranzo:2016b} and occasionally puzzles of pieces assembled from the most distant origins \cite{moreira:2008}. The identification of common elements with a shared phylogenetic history in large viral groups is relatively limited, with exceptions \cite{wolf:2018}. Recent studies suggest that a hierarchical taxonomy of large viral groups is possible \cite{koonin:2020} though, more often, viral phylogenies are limited to viral cohorts \cite{dion:2020}. The recognition of a phylogenetic signal speaks for the evolutionary continuity of the shared element, but does not inform on the actual composition of viral genomes or on the viral phenotype: the former are dominated by extensive HGT and divergent evolution, the latter by the viral ecological niche, both being interdependent. Actually, it is important to distinguish between {\it bona fide} phylogenetic elements and genes that have been recently acquired by HGT from contemporary bacteria or from the viral host \cite{moreira:2008} which, instead of revealing common origin, might represent examples of recent, likely fast, viral adaptation to new niches.

Modularity is a driving force of viral genome evolution \cite{jachiet:2014} and, probably, also a mechanism for the generation of new viral species. Actually, the proneness of viruses to loss, gain, and exchange genes has prompted the representation of large viral groups as bipartite networks with viral genomes and viral genes as the two classes of nodes \cite{iranzo:2016b,iranzo:2016c}. Each genome has as many links as genes it contains, while genes are linked to all of the genomes where they are found. From an evolutionary perspective, genes can be in principle classified into different groups, as signature genes (characteristic of one particular group of viruses), hallmark genes (encoding key proteins and shared by overlapping sets of diverse viruses), or orphan genes (found in a single genome) \cite{iranzo:2016b}. However, the bipartite network representation or viral genes and genomes allows the use of multiple quantitative measures that, in the framework of complex networks \cite{boccaletti:2006}, might both reveal the intimate architecture of gene sharing and point at dominant evolutionary mechanisms. For example, the observed scale-free distribution of the number of genomes that contain a given gene suggests a unique underlying generating mechanism. Community analyses \cite{guimera:2007} of the bipartite network corresponding to the double-stranded DNA viruses reveal, however, the existence of non-trivial correlations among genomes that translate into the consistent identification of major viral groups \cite{iranzo:2016b}. 

Our current knowledge of the virosphere is, as of yet, poor and biased \cite{tisza:2020}, a fact that certainly limits our understanding of its role as source of functional diversity. The apparent phylogenetic discontinuity between viral groups that we observe may also result from that poor sampling: a more comprehensive knowledge of viral diversity could bring about a more parsimonious understanding of how viral evolution has unfolded \cite{zhang:2018}. With obvious differences, this suggestion recalls the interpretative difficulties (and even the disdain) that surrounded paleontology, due to the incompleteness of the fossil record, until well into the 20th century. A history of the world imperfectly kept, in Darwin's words \cite{darwin:1859}, severely delayed the incorporation of paleontology as a discipline proper of evolutionary biology \cite{sepkoski:2012}. This is certainly not the case of virology, but suggests that evolutionary principles inferred from too sparse data might require substantial revision when data becomes more abundant and complete. Evolutionary theory is on the move. 

\section{Flexible cooperation and multipartite virus}

The viral world harbors amazing examples of competition and cooperation, both among kin and non-kin. Quasispecies are ensembles of genetically related genomes where multiple associations and interactions are possible, and whose composition depends on the joint action of endogenous antagonistic interactions \cite{arbiza:2010,andino:2015}. Replication at high mutation rates favors the generation of diversity, thus facilitating, in principle, adaptation \cite{sanjuan:2021}. However, too high a mutation rate might hinder the fixation of beneficial mutations \cite{stich:2007} and produce an excess of defective genomes. Defective interfering particles were originally considered as artifacts of {\it in vitro} evolution with a detrimental effect on viral fitness \cite{huang:1973}. But these particles are produced {\it in vivo} and play a role, among others, in viral adaptation and in disease progression \cite{rezelj:2018}. High mutation rates also permit the coincident appearance of mutations with similar beneficial effect in independent genomes, causing clonal interference and potentially delaying adaptation \cite{miralles:1999}. The previous effects notwithstanding, the mutation rate is itself subject to selection along evolutionary time, so it is sensible to assume that it has been tuned to favor viral survivability \cite{peck:2018}. 

Quasispecies diversity is actually needed to maintain specific viral phenotypes which, in agreement with conceptual hypotheses  \cite{eigen:1993}, are a collective property of the ensemble. A decrease in quasispecies diversity limits its adaptive ability and attenuates its pathogenic potential \cite{vignuzzi:2006}, and hinders the production of new phenotypes through cooperative interactions \cite{shirogane:2019}. Even within a population, therefore, different genotypes interact non-linearly: the whole is more than the sum of its parts. A quasispecies bears an enormous innovative potential that can be explored through genotype combinatorics. Population bottlenecks, which are common in viral propagation and facilitate the fixation of mutants \cite{escarmis:2006}, could act as filters to explore many random combinations of few genotypes simultaneously, thus benefiting viral phenotypic innovation and, eventually, viral persistence. 

Multipartite viruses take flexible cooperation to the extreme. These viruses have their genomes fragmented in a variable number of pieces, from two to eleven, that are encapsidated and propagated in independent particles \cite{sicard:2016,lucia-sanz:2017}. This lifestyle faces the risk of loosing genomic information due to the seemingly small number of viral particles that are transmitted from host to host. Until now, the advantages of such genomic organization remain unclear \cite{lucia-sanz:2018,zwart:2021}, though there are two important factors that may have contributed to the repeated emergence of multipartite viruses in evolution: the possibility to adapt to new hosts through gene copy number variation \cite{sicard:2013} and the advantage conferred by fast adaptation through rapid associations with non-kin when new niches become available \cite{lucia-sanz:2017}.
The scenario where multipartitism emerges as a successful adaptive strategy of the type first-come first-served is supported by a number of empirical observations. First, there was a fast radiation of multipartite viruses when agriculture became common practice \cite{gibbs:2010}; second, the genome of some multipartite viruses has genes originated in different viral families \cite{koonin:2015b}; third, many viruses undergo transient associations with subviral particles, such as virus satellites, that change the viral phenotype \cite{krupovic:2016}. This flexible cooperation might be a first step to permanent associations in the form of a bipartite virus: there are examples of viral families with virus-satellite associations and bipartite species, as Geminiviruses  \cite{shah:2009}. Fourth, multipartite viruses rapidly modify the copy number of each genomic fragment from one host species to another \cite{michalakis:2020}.

Multipartite viruses infect mostly plants, which are often simultaneously infected by viruses of different families \cite{elena:2014}. This permissiveness may underlie the exploration of new associations among viral genes. The route to multipartitism, however, should not be unique as, beyond {\it de novo} associations, gene duplication or genome fragmentation may also promote the emergence of multipartite viral species \cite{lucia-sanz:2017}. Multipartitism has appeared multiple times in evolution and, as such, it has to be understood as a successful evolutionary strategy \cite{maynard-smith:1982}.

\section{Conclusion}
The genotype-to-function map is many-to-many. Many genotypes can code for similar phenotypes and each genotype has the potential to express a variety of phenotypes. As a consequence, the map is proto-functional and highly adaptable. The exploration of evolutionary innovations, further, is a process that runs in parallel under many different selective conditions. Also, what is not useful in a certain context  may provide an advantage and thrive in another. HGT in its many expressions promotes this distributed assay-select-share-combine process. Once a variety of pieces is in place, spontaneous cooperative associations can give rise to new levels of complexity. 

Evolution operates at such long time scales that even detailed observations of the here and now turn out to be insufficient to educate our intuition on the diversity of complex molecular organizations possible, and on the underlying mechanisms. It happens often that certain evolutionary pathways are disregarded only because we never considered them as a possibility or never looked for their products. An example is the idea of a {\it de novo} generation of function: only in the last decade have we uncovered how genes can be generated from non-genic sequences \cite{carvunis:2012,vakirlis:2020}, how transposable elements can be "domesticated" to perform specific functions in their host \cite{sinzelle:2009}, how promoters emerge from random sequences \cite{yona:2018} and how this can happen even in the absence of sequence diversity, simply through successive cycles of mutation, enrichment and selection \cite{wachowius:2021}. 

Evolution is a powerful tinkerer. It will use any mechanism that is available, low-cost, and constructive in a very generic way. It uses from phenotypic redundancy to gene combinatorics. When all these elements are taken into consideration, what comes as a difficulty is to imagine a world devoid of molecular complexity. 

\subsection*{Acknowledgments}
{The author is grateful to José A. Cuesta, Ester Lázaro, and Luis F. Seoane for their insightful comments. This work has been funded by the Spanish Ministerio de Ciencia, Innovaci\'on y Universi\-da\-des-FEDER funds of the European Union support, under project MiMevo (FIS2017-89773-P). The Spanish MICINN has also funded the ``Severo Ochoa'' Centers of Excellence to CNB, SEV 2017-0712.}

\bibliography{bibliography}

\end{document}